\documentstyle[aps,prd]{revtex}
\tighten
\renewcommand{\epsilon}{\varepsilon}

\begin{document}

\title{Cosmological particle production and generalized 
thermodynamic equilibrium}

\author{Winfried Zimdahl\footnote{Electronic address: winfried.zimdahl@uni-konstanz.de}}
\address{Fakult\"at f\"ur Physik, Universit\"at Konstanz, PF 5560 M678
D-78457 Konstanz, Germany\footnote{present address}
\\
Departament de F\'{\i}sica,
Universitat Aut\`{o}noma de Barcelona,
E-08193 Bellaterra (Barcelona), Spain
\\
and 
Laboratoire de Gravitation et Cosmologie Relativistes,  
Universit\'{e} Pierre et Marie Curie- CNRS/URA 769 \\
Tour 22, 4\`{e}me \'{e}tage, Bo\^{\i}te 142, 
4, place Jussieu, 75252 Paris Cedex 05, France}

\date{\today}

\maketitle


\begin{abstract}
With the help of a conformal, timelike Killing-vector we define 
generalized equilibrium states for cosmological fluids with particle production. 
For massless particles the generalized 
equilibrium conditions require 
the production rate to 
vanish and 
the well known ``global'' equilibrium 
of standard relativistic thermodynamics is recovered as a limiting case. 
The equivalence between the creation rate for particles with 
nonzero mass and 
an effective viscous fluid pressure follows as a consequence of the  generalized equilibrium properties. 
The implications of this equivalence for the cosmological dynamics are 
discussed, including the possibility of a power-law inflationary behaviour. 
For a simple gas a microscopic derivation for such kind of 
equilibrium is given on the basis of relativistic kinetic theory.

\end{abstract}

\section{Introduction}

Simple relativistic gases or fluids may be in a state of ``global''
thermodynamic equilibrium if the spacetime admits a 
timelike Killing-vector (KV) i.e., if it is stationary 
\cite{TauWei,Cher,WI,Stew,Ehl,IS,Neugeb,Groot,Bern,Maar}. 
The immediate consequence of this well-known fact is that such kind of equilibrium cannot generally exist in the 
homogeneous and isotropic Friedmann-Lema\^{\i}tre-Robertson-Walker (FLRW)  models of the expanding 
Universe since these models are not stationary. 
The only exception is a system of relativistic, massless particles 
(radiation), for which the ``global'' equilibrium condition is less 
restrictive and requires 
the existence of a conformal timelike Killing-vector only. 
A conformal timelike Killing-vector however, is admitted in FLRW 
spacetimes. 

As was pointed out recently \cite{TZP,ZTP}, this general picture representing 
common wisdom is modified if one drops the assumption of particle number conservation which is always implied in the mentioned standard considerations. 
If the number of simple gas particles is increasing, i.e., 
if there is net particle production (of whatsoever origin), there exist collisional equilibrium states 
which we will call here ``generalized'' equilibrium states,  
relying on the existence of a 
conformal Killing-vector (CKV), for {\it any} equation 
of state 
admitted by the standard kinetic theory of a simple relativistic gas, not 
only for radiation. 
Radiation under these conditions is the limiting case of 
vanishing particle production. 
Denoting the fluid energy density by $\rho $, the equilibrium pressure 
by $p$, 
the temperature by $T$ and the fluid four-velocity by $u ^{i}$, it was 
shown that 
for any equation of state within the range between nonrelativistic matter 
($p \ll \rho$) and radiation ($p = \rho /3$) there exists a well-defined 
rate at which the corresponding particles have to be produced in order to satisfy the (generalized) equilibrium conditions following from the CKV property of $u _{i}/T$. 
This rate is highest for nonrelativistic matter ($p \ll \rho $) in which 
case it amounts to half the expansion rate and it vanishes, as already mentioned, in the opposite limit of radiation ( $p = \rho /3$). 

All results concerning this kind of ``generalized '' 
equilibrium states characterized by a 
timelike CKV under the condition of increasing particle number were 
obtained within a simple particle creation model on the level of 
relativistic kinetic theory. 
A modified Boltzmann equation was proposed in which a 
``source term'' describes the change of the one-particle distribution 
function due to particle number nonconserving processes, supposedly of 
quantum origin. 
The above statements on the creation rates necessary for the generalized 
equilibrium conditions to be fulfilled were then obtained from a simple rate 
approximation of the source term in Boltzmann's equation. 
Therefore the question arises, whether the new equilibrium states 
associated with a varying particle number are just features of a 
specific creation model or whether they reflect a more general 
property of corresponding physical systems. 

The main purpose of the present paper is to clarify this situation. 
It will turn out that the characterization of generalized thermodynamic 
equilibrium through a conformal, timelike KV for a fluid with 
nonconserved particle number as sketched above is completely general 
and does not depend at all on a specific creation model. 
This conclusion is obtained on a purely phenomenological basis without 
any assumptions other than standard ones in fluid cosmology, except for admitting a nonvanishing source term in the particle number balance. 

Furthermore, the equivalence between a nonvanishing cosmological particle production rate and an effective bulk pressure is shown to be a 
consequence 
of the generalized equilibrium conditions provided standard assumptions for  imperfect fluids are made. 
Especially, it will not be necessary to introduce a ``creation'' 
pressure by hand. 
This supplementary pressure may be derived and needs 
not to be postulated. 

The second purpose of the paper is to overcome some limitations of the previously used kinetic model for particle production \cite{TZP,ZTP}. 
Within the latter model the equivalence between production rate and 
effective bulk pressure was restricted to spatially homogeneous spacetimes. 
We show here that this restriction is unnecessary. 
We will establish an exact correspondence between the kinetic theory model 
for particle production and the general, model independent phenomenological approach. 

The plan of the paper is as follows. 
In Sec. II we define a phenomenological concept of 
generalized equilibrium which is compatible with an increase in the fluid particle number. 
Section III investigates the thermodynamics of the corresponding creation process with special emphasis on the limiting cases for pure radiation and nonrelativistic matter. 
Section IV considers the cosmological dynamics for a universe filled with 
matter in generalized equilibrium. 
Section V presents a microscopic justification of the generalized equilibrium concept for a relativistic gas. 
A brief summary of the paper is given in Sec. VI.  
Units have been chosen so that $c = k_{B} = \hbar = 1$. 

\section{Generalized equilibrium}

The energy-momentum tensor of an imperfect fluid is generally given by 
(see, e.g., \cite{Groot})
\begin{equation}
T ^{ik} = T ^{ik}_{\left(0 \right)} + \pi h ^{ik} 
+ \pi ^{ik} + q ^{i}u ^{k} + q ^{k}u ^{i}
\label{1}
\end{equation}
with 
\begin{equation}
T ^{ik}_{\left(0 \right)} = \rho  u ^{i}u ^{k} + p h ^{ik}
\label{2}
\end{equation}
and 
\begin{equation}
\pi ^{ik}u _{k} = q ^{i}u _{i} = 
\pi ^{i}_{i} =  h ^{ik}u _{i} = 0\ ,
\ \ \ \ \ \ u ^{i}u _{i} = -1 \ . 
\label{3}
\end{equation}
Here, $\rho $ is the energy density of a fiducial equilibrium state, 
$p$ is the corresponding equilibrium pressure, $u ^{i}$ is the fluid four-velocity in the Eckart frame and $h ^{ik} = g ^{ik} + u ^{i}u ^{k}$ 
is the spatial projection tensor. 
The quantity $\pi $ denotes that part of the scalar pressure which is 
connected with entropy production, $- \pi ^{ik}$ is the anisotropic 
stress tensor and 
$q ^{i}$ is the heat flux vector. 
Within the Eckart frame the particle number flow vector $N ^{i}$ is 
given by 
\begin{equation}
N ^{i} = n u ^{i}
\label{4}
\end{equation}
where $n$ is the particle number density. 
While we require local energy-momentum conservation $T ^{ik}_{\ ;k} = 0$ 
in accordance with the integrability conditions of Einstein's equations, 
we will not assume the fluid particle number to be conserved. Instead, 
we admit a source term $\Gamma $ in the corresponding balance equation 
which describes the rate of change of the fluid particles:
\begin{equation}
N ^{i}_{;i} = n \Gamma \ .
\label{5}
\end{equation}
Inserting here the expression (\ref{4}) yields
\begin{equation}
\dot{n} + \Theta n = n \Gamma 
\label{6}
\end{equation}
where $\Theta \equiv  u ^{i}_{;i}$ is the fluid expansion and 
$\dot{n} \equiv  n _{,i}u ^{i}$. 
The quantity $\Gamma $ is assumed to describe the net change of the 
particle number either due to internal reactions within the medium \cite{ZMNRAS,ZPM} or due to particle production in strongly varying gravitational fields 
\cite{Zel,Mur,Hu,BiDa,Ford}. 
Usually, it is an input quantity in a phenomenological description which 
for specific processes has 
to be calculated from the underlying microphysics. 
In our considerations $\Gamma $ will be fixed by the generalized equilibrium conditions to be discussed below.  

Projecting the covariant derivative of $T ^{ik}$ in direction of $u _{i}$ 
provides us with  
\begin{equation}
u _{i}T ^{ik}_{;k} = u _{i}T ^{ik}_{\left(0 \right);k} 
+ u _{i}\left[\pi h ^{ik} 
+ \pi ^{ik} + q ^{i}u ^{k} + q ^{k}u ^{i} \right]_{;k} = 0 \ .
\label{7}
\end{equation}
The first term on the right-hand side of the last equation may be written 
as
\begin{equation}
u _{i}T ^{ik}_{\left(0 \right);k} = - \dot{\rho } 
- \left(\rho + p \right)\Theta \ .
\label{8}
\end{equation}
From the Gibbs equation 
\begin{equation}
T \mbox{d}s = \mbox{d} \frac{\rho }{n} + p \mbox{d}\frac{1}{n}\ ,
\label{9}
\end{equation}
where $s$ is the entropy per particle, we obtain 
\begin{equation}
\dot{\rho } = n T \dot{s} + \left(\rho + p \right)
\frac{\dot{n}}{n}\ .
\label{10}
\end{equation}
Using here  
the particle number balance (\ref{6}) yields
\begin{equation}
u _{i}T ^{ik}_{\left(0 \right);k} = - n T \dot{s} 
- \left(\rho + p \right)\Gamma \ .
\label{11}
\end{equation}
Inserting the last relation into the balance (\ref{7}) the latter becomes 
\begin{equation}
n \dot{s} + \left(\frac{q ^{k}}{T} \right)_{;k} 
= - \frac{\rho + p}{T}\Gamma 
- \frac{1}{2}\left(T ^{ik} - T ^{ik}_{\left(0 \right)} \right)
\left[\left(\frac{u _{i}}{T}\right)_{;k} 
+ \left(\frac{u _{k}}{T}\right)_{;i}\right]\ .
\label{12}
\end{equation}
Taking into account the standard definition 
\begin{equation}
S ^{i} = n s u ^{i} + \frac{q ^{i}}{T}
\label{13}
\end{equation}
of the entropy flow vector $S ^{i}$ of the Eckart theory 
allows us to write Eq. (\ref{12}) as
\begin{eqnarray}
S ^{i}_{;i} - n s \Gamma  &=&  
- \frac{\rho + p}{T}\Gamma 
- \frac{1}{2}\left(T ^{ik} - T ^{ik}_{\left(0 \right)} \right)
\left[\left(\frac{u _{i}}{T}\right)_{;k} 
+ \left(\frac{u _{k}}{T}\right)_{;i}\right]        \nonumber\\
&=& - \frac{\rho + p}{T}\Gamma 
- \frac{1}{2}\left(T ^{ik} - T ^{ik}_{\left(0 \right)} \right)
\pounds _{u _{i}/T} g _{ik}\ ,
\label{14}
\end{eqnarray}
where $\pounds _{u _{i}/T}$ denotes the Lie derivative with respect to the temperature vector 
$u _{a}/T$. 
For $\Gamma = 0$ equation (\ref{14}) reduces to the well-known expression 
for the entropy production density of a fluid with constant 
particle number  \cite{Neugeb,Steph}. 
In the latter case one has vanishing entropy production for 
$\pounds _{u _{i}/T} g _{ik} = 0$, i.e., if $u _{a}/T$  is a KV. 
This statement has a counterpart on the level of kinetic theory following 
from the requirement that the equilibrium distribution function for a simple gas obeys Boltzmann's equation (see, e.g, \cite{Stew,Ehl,IS} and below). 

For $\Gamma \neq 0$, however, there exists the following possibility to characterize a different kind of equilibrium states in case $u _{a}/T$ 
is a CKV, i.e., if 
\begin{equation}
\left(\frac{u _{i}}{T} \right)_{;k} 
+ \left(\frac{u _{k}}{T} \right)_{;i} 
= 2 \phi g _{ik}
\label{15}
\end{equation}
is fulfilled. 
Eq. (\ref{14}) then becomes 
\begin{equation}
S ^{i}_{;i} - ns \Gamma 
= - \frac{\rho + p}{T}\Gamma - 3 \pi \phi \ .
\label{16}
\end{equation}
Since the CKV property implies the relation 
\begin{equation}
\phi = \frac{1}{3}\frac{\Theta }{T}\ ,
\label{17}
\end{equation}
the right-hand side of Eq. (\ref{16}) vanishes under the condition 
\begin{equation}
\pi = - \left(\rho + p \right)\frac{\Gamma }{\Theta }\ .
\label{18}
\end{equation}
The entropy production density in such a case is given by
\begin{equation}
S ^{i}_{;i} = ns \Gamma \ ,
\label{19}
\end{equation}
where $\Gamma \geq 0 $, i.e., production of particles is required to satisfy the second law of thermodynamics. 
The term $n s \Gamma $ on the right-hand side of Eq. (\ref{19}) describes the increase of the entropy of the system due to the fact that its phase space 
is enlarged. 
Each of the created particles carries the same entropy 
$s$ as the particles already around.   
From the moment of its creation each particle contributes the equilibrium amount $s$ to the entropy of the system. 
This is equivalent to the statement that the produced particles share the equilibrium properties of the already existing fluid particles from the very beginning. 
It is only the amount of entropy describing the very existence of additional fluid particles at equilibrium with the particles already there which counts 
on the right-hand side of Eq. (\ref{19}). 
In a microscopic picture, all particles including the just produced ones are governed by the same equilibrium distribution function (see Eq. (\ref{73}) below). 
Especially, no terms describing 
dissipative processes such as viscosities or heat fluxes 
appear on the right-hand side of Eq. (\ref{19}). 
This justifies the classification of the relationship (\ref{19}) as an equilibrium situation although there is an increase in the entropy of the system. 
{\it We define the ``generalized equilibrium''  
by (i) the vanishing of 
$S ^{a}_{;a} - n s \Gamma $, i.e., by Eq. (\ref{19}) with 
$\Gamma \geq 0$ together with (ii) the CKV property (\ref{15}) 
of $u _{i}/T$}. 
For $\Gamma = 0$ this requirement reduces to the condition $\pi = 0$ 
(cf. Eq. (\ref{18})), a case dealt with by Bedran and Calv\~{a}o \cite{BedCalv}. 
Equation (\ref{18}) for $\Gamma \neq 0$, however, coincides with the 
frequently used condition for ``adiabatic particle production''. 
This condition is a consequence of 
the definition (\ref{13}) and the generalized equilibrium requirements. 
There is no need to postulate the existence of a supplementary ``creation pressure'' separately. 
The existence of such kind of pressure is a consequence 
of the equilibrium conditions and needs not to be assumed. 

To obtain an alternative expression for the entropy production density we realize that differentiating the expression (\ref{13}) yields 
\begin{equation}
S ^{i}_{;i} - n s \Gamma = n \dot{s} + \frac{\nabla _{a}q ^{a}}{T} 
+ \frac{q ^{a}}{T}\left(\frac{\nabla _{a}T}{T} + 
\dot{u}_{a} \right) \ ,
\label{20}
\end{equation}
where $\nabla _{a}q ^{a} \equiv  h _{a}^{b}q ^{a}_{;b}$ etc. 
On the other hand, the energy balance (\ref{7}) may be written as 
\begin{equation}
\dot{\rho } = - \Theta \left(\rho + p + \pi  \right) 
- \nabla _{a}q ^{a} - 2 \dot{u}_{a}q ^{a} 
- \sigma _{ab}\pi ^{ab}
\label{21}
\end{equation}
with
\begin{equation}
\sigma _{ab} = \frac{1}{2}\left(\nabla _{a}u _{b} 
+ \nabla  _{b}u _{a} - \frac{2}{3}h _{ab}\Theta \right). 
\label{22}
\end{equation}
Consequently, via Eqs. (\ref{10}) and (\ref{6}) the quantity $\dot{s}$ 
is given by 
\begin{equation}
n T \dot{s} = - \Theta \pi - \left(\rho + p \right)\Gamma 
- \nabla  _{a}q ^{a} - 2 \dot{u}_{a}q ^{a} 
- \sigma _{ab}\pi ^{ab}\ .
\label{23}
\end{equation}
Combining relations (\ref{20}) and (\ref{23}) we obtain
\begin{equation}
S ^{i}_{;i} - n s \Gamma = - \frac{\Theta \pi}{T} 
- \frac{\rho + p}{T}\Gamma 
- \frac{q ^{a}}{T}\left(\frac{\nabla  _{a}T}{T} + \dot{u}_{a} \right) 
- \frac{\sigma _{ab}\pi ^{ab}}{T}
\label{24}
\end{equation}
which is an alternative way of writing Eq. (\ref{14}). 
Since the CKV conditions (\ref{15}) imply 
\begin{equation}
\frac{\nabla  _{a}T}{T} + \dot{u}_{a} = \sigma _{ab} = 0 \ ,
\label{25}
\end{equation}
the right-hand side of Eq. (\ref{24}) vanishes, provided condition 
(\ref{18}) 
is satisfied. 
It may be worth mentioning that condition (\ref{18}) is not generally equivalent 
to $\dot{s} = 0$ but only if the heat flux $q ^{a}$ on the right-hand side 
of Eq. (\ref{23}) does not contribute. 
An important consequence of the CKV property (\ref{15}) is that it 
completely fixes the evolution of the temperature:
\begin{equation}
\frac{\dot{T}}{T} = - \frac{\Theta }{3}\ .
\label{26}
\end{equation}
On the other hand, the behaviour of the temperature is determined by 
general thermodynamical considerations. 
The fluid equations of state may be written as 
\begin{equation}
p = p \left(n,T \right)\ ,\ \ \ \ \ \ \ \ \ 
\rho = \rho \left(n,T \right)\ .
\label{27}
\end{equation}
Differentiating the latter relation and using 
the balances (\ref{6}) and  
(\ref{10}) we obtain
\begin{equation}
\frac{\dot{T}}{T} = - \left(\Theta - \Gamma  \right) 
\frac{\partial p}{\partial \rho } 
+ \frac{n \dot{s}}{\partial \rho / \partial T}\ ,
\label{28}
\end{equation}
where the abbreviations
\[
\frac{\partial{p}}{\partial{\rho }} \equiv  
\frac{\left(\partial p/ \partial T \right)_{n}}
{\left(\partial \rho / \partial T \right)_{n}} \ ,
\ \ \ \ \ \ \ \ 
\frac{\partial{\rho }}{\partial{T}} \equiv  
\left(\frac{\partial \rho }{\partial T} \right)_{n}
\]
have been used. 
Comparing Eqs. (\ref{26}) and (\ref{28}) and assuming fom now on 
$q ^{a} = 0$, i.e. $\dot{s} = 0$, 
yields 
\begin{equation}
\Gamma = \left(1 - \frac{1}{3}\frac{\partial \rho }
{\partial p} \right)\Theta \ .
\label{29}
\end{equation}
The CKV property fixes the creation rate. 
An equilibrium according to $S ^{a}_{;a} - ns \Gamma = 0$ (generalized equilibrium) is only possible 
if fluid particles are created at a specific rate given by Eq. (\ref{29}). 
The production rate $\Gamma $ is positive for any equation of state 
$0 < p \leq \rho /3$. 
It vanishes for $p = \rho /3$, i.e. for 
radiation. 
This means recovering the well-known fact that a fluid obeying the 
equation of state $p = \rho /3$ may be in ``global equilibrium'' if the spacetime admits a 
timelike CKV. 
The CKV property of $u _{i}/T$ implies the temperature law (\ref{26}) 
which just coincides with the well-known temperature behaviour 
(adiabatic cooling) of a relativistic fluid in the expanding universe. 
For $\Gamma = 0$ only a fluid with the equation of state $p = \rho /3$ 
satisfies Eq. (\ref{26}).   
For any other equation of state the equilibrium conditions require 
$\phi = 0$ in Eq. (\ref{15}), i.e., the existence of a KV. 
If one drops the condition $\Gamma = 0$, i.e., if one allows the particle number to increase, radiation is no longer a singular case. 
Generalized equilibrium states (\ref{15}) with $\phi \neq 0$ 
are possible for any equation of state in the range 
$0 < p \leq \rho /3$.   
For any of these equation of state there exists a specific production 
rate (\ref{29}) which is necessary to satisfy the 
generalized equilibrium conditions (\ref{15}) and (\ref{19}). 
The production of particles at a rate different from Eq. (\ref{29}) disturbs the generalized equilibrium.  
In other words, a fluid with an equation of state in the range 
$0 < p < \rho /3$ may only be in generalized 
equilibrium if fluid particles are produced 
at a specific rate depending on the equation of state. 
A specific rate $\Gamma $ on the other hand, uniquely fixes $\pi $ acording 
to condition (\ref{18}):
\begin{equation}
\pi = - \left(\rho + p \right)
\left(1 - \frac{1}{3}\frac{\partial \rho }
{\partial p} \right)\ .
\label{30}
\end{equation}
We recall that the 
generalized equilibrium conditions naturally imply the existence of 
an effective bulk pressure, i.e., the latter has not to be postulated separately. 
It is obvious that $\pi $ as well as $\Gamma $ vanish for $p = \rho /3$, 
i.e. for radiation. 
For nonrelativistic matter, characterized by equations of state 
$p = nT$ and $\rho = nm + \frac{3}{2}nT$ with $m \gg T$ one obtains 
\begin{equation}
\Gamma = \frac{1}{2}\Theta \ ,
\ \ \ \ \ \ \ \ \  
\pi = - \frac{1}{2}\left(\rho + p \right) 
\approx - \frac{\rho }{2}\ ,
\ \ \ \ \ \ \ \ \ (m \gg T)\ ,
\label{31}
\end{equation}
i.e., $\Gamma $ is half the expansion rate and $|\pi|$ is a remarkable 
fraction of the energy density. 

\section{Thermodynamics of the creation process}
With equations of state of the general form (\ref{27}) the particle number density $n$ and the temperature $T$ are the primary thermodynamic variables 
of our system. 
While the (generalized) equilibrium behaviour of the temperature is given by 
Eq. (\ref{26}) independently of the equation of state, the particle number density changes according to 
\begin{equation}
\frac{\dot{n}}{n} = - \frac{1}{3} \frac{\partial{\rho }}{\partial{p}} 
\Theta \ ,
\label{32}
\end{equation}
where we have combined Eqs. (\ref{6}) and (\ref{29}). 
Use of Eq. (\ref{30}) in Eq. (\ref{21}) yields 
(recall that $q ^{a} = \sigma ^{ab} = 0$) 
\begin{equation}
\dot{\rho } = - \frac{\Theta }{3}\left(\rho + p \right)
\frac{\partial{\rho }}{\partial{p}} \ .
\label{33}
\end{equation}
Differentiating the first equation of state (\ref{27}) and applying 
relations (\ref{6}) and (\ref{28}) under the condition $\dot{s} = 0$, i.e., 
with the equations (\ref{18}) and (\ref{29}), 
one finds
\begin{equation}
\dot{p} = c _{s}^{2}   \dot{\rho }
\label{34}
\end{equation}
where 
\begin{equation}
c _{s}^{2}  = \left(\frac{\partial{p}}{\partial{\rho }} \right)_{ad} 
= \frac{n}{\rho + p}\frac{\partial{p}}{\partial{n}} 
+ \frac{T}{\rho + p} 
\frac{\left(\partial p / \partial T \right)^{2}}
{\partial \rho / \partial T}
\label{35}
\end{equation}
is the square of the adiabatic sound velocity. 
An essential quantity to characterize the particle production process is 
$\mu /T$, where $\mu $ is the chemical potential. 
Using Eqs. (\ref{26}) and (\ref{34}) in the Gibbs-Duhem relation (see, e.g., \cite{Groot})
\begin{equation}
\mbox{d} p = \left(\rho + p \right)\frac{\mbox{d} T}{T} 
+ n T \mbox{d} \left(\frac{\mu }{T} \right)
\label{36}
\end{equation}
provides us with 
\begin{equation}
\left(\frac{\mu }{T} \right)^{^{\displaystyle \cdot}} = 
\frac{\Theta }{3}\frac{\rho + p}{n T}
\left[1 - c _{s}^{2}  \frac{\partial{\rho }}{\partial{p}}\right] 
= \left[\frac{\rho }{p} - \frac{\partial{\rho }}{\partial{p}} \right]
\frac{\Theta }{3}\ .
\label{37}
\end{equation}
Here we have used that for $p =nT$, i.e. for a classical gas, 
the square of the sound velocity 
(\ref{35}) may be written as 
\begin{equation}
c _{s}^{2}  = \frac{nT}{\rho + p}
\left(1 + \frac{\partial{p}}{\partial{\rho }} \right)\ .
\label{38}
\end{equation}
The right-hand side of Eq. (\ref{37}) vanishes only for radiation 
($c _{s}^{2}  = 1/3$ and $\partial \rho / \partial p = 3$). 
Characterizing the equilibrium states under consideration also requires 
information about the spatial derivative of $\mu /T$. 
From the momentum balance 
\begin{equation}
\left(\rho + p + \pi  \right)\dot{u} _{m} 
+ \nabla  _{m}\left(p + \pi  \right) = 0
\label{39}
\end{equation}
(where we assumed $q ^{a} = \pi ^{ab} = 0$) 
and the Gibbs-Duhem equation (\ref{36}) together with Eq. 
(\ref{25}) we obtain 
\begin{equation}
n T \nabla  _{a}\left(\frac{\mu }{T} \right) = 
- \pi \dot{u}_{a} - \nabla  _{a}\pi \ ,
\label{40}
\end{equation}
where $\pi $ is given by Eq. (\ref{30}). 
It follows that also the spatial derivative of $\mu /T$ vanishes for 
radiation. 
Restricting ourselves to a classical gas and 
using  Eqs. (\ref{30}) 
and 
(\ref{25}), Eq. (\ref{40}) may be written as 
\begin{equation}
\frac{\nabla  _{a}\pi }{\rho + p} = 
- \left(1 - \frac{1}{3}\frac{\partial{\rho }}{\partial{p}} \right)
\frac{\nabla  _{a}T}{T} 
- \frac{n T}{\rho + p}\nabla  _{a}\left(\frac{\mu }{T} \right)\ .
\label{41}
\end{equation}
On the other hand, an alternative expresssion for $\nabla  _{a}\pi $ 
in terms of $\nabla  _{a}T$ and $\nabla  _{a}\mu $ may be obtained 
directly from 
Eq. (\ref{30}). 
To this purpose we first realize that the Gibbs-Duhem equation (\ref{36}) implies
\begin{equation}
\frac{\nabla  _{a}p}{\rho + p} = \frac{\nabla  _{a}T}{T} 
+ \frac{n T}{\rho + p}\nabla  _{a}\left(\frac{\mu }{T} \right)\ .
\label{42}
\end{equation}
With $p = nT$ we obtain 
\begin{equation}
\frac{\nabla  _{a}n}{n} = \nabla  _{a}\left(\frac{\mu }{T} \right) 
+ \frac{\rho }{p}\frac{\nabla  _{a}T}{T}\ .
\label{43}
\end{equation}
From the Gibbs equation (\ref{9}) it follows that 
\begin{equation}
\frac{\nabla  _{a }\rho }{\rho + p} = \frac{nT}{\rho + p}\nabla  _{a}s 
+ \frac{\nabla  _{a}n}{n}\ .
\label{44}
\end{equation}
Spatial differentiation of  
the second equation of state (\ref{27}) provides us with 
\begin{equation}
\frac{\nabla  _{a}\rho }{\rho + p} = 
\left(1 - \frac{T}{\rho + p}\frac{\partial{p}}{\partial{T}} \right)
\frac{\nabla  _{a}n}{n} 
+ \frac{T}{\rho + p}\frac{\partial{\rho }}{\partial{T}}
\frac{\nabla  _{a}T}{T}\ ,
\label{45}
\end{equation}
where we have used the equation 
\begin{equation}
\frac{\partial{\rho }}{\partial{n}} = \frac{\rho + p}{n} 
- \frac{T}{n}\frac{\partial{p}}{\partial{T}}\ ,
\label{46}
\end{equation}
which follows from the fact that the entropy is a state function. 
Comparing the expressions (\ref{44}) and (\ref{45}) and using 
Eq. (\ref{43}) yields 
\begin{equation}
\nabla  _{a}\left(s + \frac{\mu }{T} \right) = 
\left[\frac{\partial{\rho }}{\partial{p}} - \frac{\rho }{p}\right]
\frac{\nabla  _{a}T}{T}\ .
\label{47}
\end{equation}
Realizing now that \cite{Groot} 
\begin{equation}
s + \frac{\mu }{T} = \frac{\rho + p}{nT}\ ,
\label{48}
\end{equation}
use of $p = nT$ and Eqs. (\ref{43}) and (\ref{47}) yields for the spatial gradient 
of the effective viscous pressure $\pi $ in 
Eq. (\ref{30})
\begin{eqnarray}
\frac{\nabla  _{a}\pi }{\rho + p} &=& 
- \left(1 - \frac{1}{3}\frac{\partial{\rho }}{\partial{p}} \right)
\nabla  _{a}\left(\frac{\mu }{T}\right) 
- \left(1 - \frac{1}{3}\frac{\partial{\rho }}{\partial{p}} \right)
\frac{\rho + p}{nT}\frac{\nabla  _{a}T }{T} 
       \nonumber\\
&& - \frac{nT}{\rho + p}\left(1 - \frac{1}{3}\frac{\partial{\rho }}{\partial{p}} \right)
\left(\frac{\partial{\rho }}{\partial{p}} - \frac{\rho }{p} \right)
\frac{\nabla  _{a}T}{T} 
+ \frac{1}{3}\nabla  _{a}\left(\frac{\partial{\rho }}{\partial{p}} \right)
\ .
\label{49}
\end{eqnarray}
For a gas one has \cite{ZTP}
\begin{equation}
\frac{\partial{\rho }}{\partial{p}} = \left(\frac{m}{T} \right)^{2} 
- 1 + 5 \frac{\rho + p}{n T} 
- \left(\frac{\rho + p}{n T} \right)^{2}\ .
\label{50}
\end{equation}
Consequently, the expression (\ref{49}) for $\nabla  _{a}\pi $ becomes, in terms of $\nabla  _{a}T$ and $\nabla  _{a}(\mu /T)$,
\begin{eqnarray}
\frac{\nabla  _{a}\pi }{\rho + p} &=& 
- \left(1 - \frac{1}{3}\frac{\partial{\rho }}{\partial{p}} \right)
\nabla  _{a}\left( \frac{\mu }{T} \right)
- \left(1 - \frac{1}{3}\frac{\partial{\rho }}{\partial{p}} \right)
\frac{\rho + p}{nT}\frac{\nabla  _{a}T }{T} 
       \nonumber\\
&& + \left[- \frac{nT}{\rho + p}\left(1 - \frac{1}{3}\frac{\partial{\rho }}{\partial{p}} \right)
\left(\frac{\partial{\rho }}{\partial{p}} - \frac{\rho }{p} \right) 
- \frac{2}{3}\left(\frac{m}{T} \right)^{2} 
+ \frac{1}{3}\left(5 - 2 \frac{\rho + p}{nT} \right)
\left(\frac{\partial{\rho }}{\partial{p}} - \frac{\rho }{p}\right)
\right]
\frac{\nabla  _{a}T}{T} 
\ .
\label{51}
\end{eqnarray}
Comparing now the expressions (\ref{41}) and (\ref{51}) provides us with a relation for $\nabla  _{a}\left(\mu /T \right)$ in terms of $\nabla  _{a}T$: 
\begin{eqnarray}
- \left(\frac{\rho }{\rho + p} - \frac{1}{3}
\frac{\partial{\rho }}{\partial{p}} \right)
\nabla  _{a}\left(\frac{\mu }{T} \right) &=& 
\left[\left(1 - \frac{1}{3}\frac{\partial{\rho }}{\partial{p}} \right)
\left(\frac{\rho }{p} + \frac{nT}{\rho + p}
\left[\frac{\partial{\rho }}{\partial{p}} - \frac{\rho }{p} \right] \right) 
\right. \nonumber\\
&& \left. \ \ \ \ \ \ 
+ \frac{2}{3}\left(\frac{m}{T}\right)^{2} 
+ \left(\frac{2}{3}\frac{\rho }{p} - 1\right)
\left(\frac{\partial{\rho }}{\partial{p}} - \frac{\rho }{p}\right)
\right] \frac{\nabla  _{a}T}{T} \ .
\label{52}      
\end{eqnarray}
The case of vanishing parenthesis in front of 
$\nabla  _{a}\left(\mu /T \right)$ on the left-hand side of the last 
equation, i.e., 
$\rho / \left(\rho + p \right) - \left(1/3 \right)
\partial \rho / \partial p = 0$ corresponds to 
(cf. Eq. (\ref{30})) $\pi = - p 
= - nT$. 
In such a case the momentum balance (\ref{39}) reduces to 
$\dot{u}_{m} = 0$ 
which, according to Eq. (\ref{25}), implies $\nabla  _{a}T = 0$. 
Then, $\nabla  _{a}\left(\mu /T \right)$ by Eq. (\ref{43}) is 
simply given through  
\begin{equation}
\nabla  _{a}\left(\frac{\mu }{T} \right) = \frac{\nabla  _{a}n}{n}\ .
\label{53}
\end{equation}
Of special interest are again the limiting cases of radiation and nonrelativistic matter. 
In the first case, i.e., for $p = \rho /3$, Eq. (\ref{52}) yields 
\begin{equation}
\nabla  _{a}\left(\frac{\mu }{T} \right) = 0 
\mbox{\ \ \ \ \ \ \ \ \ \ \ \ } 
\left(p = \rho /3 \right) \ .
\label{54}
\end{equation}
The quantity $\mu /T$ is constant both in space and time 
(cf. Eq. (\ref{37})). 
In the opposite limit $p = nT$ and $\rho = nm + \frac{3}{2}nT$ 
with 
$T \ll m$ one obtains 
\begin{equation}
\nabla  _{a}\left(\frac{\mu }{T} \right) = - \frac{m}{T}
\frac{\nabla  _{a}T}{T}
\mbox{\ \ \ \ \ \ \ \ \ \ }
\left(m \gg T  \right) \ ,
\label{55}
\end{equation}
which may be written as
\begin{equation}
\nabla  _{a}\left(\frac{\mu - m}{T} \right) = 0 
\mbox{\ \ \ \ \ \ \ \ \ \ }
\left(m \gg T  \right) \ .
\label{56}
\end{equation}
Since one finds from relation (\ref{37}) that also 
\begin{equation}
\left(\frac{\mu - m}{T} \right)^{^{\displaystyle \cdot}} = 0 
\mbox{\ \ \ \ \ \ \ \ \ \ } 
\left(m \gg T  \right) 
\label{57}
\end{equation}
holds in this limit, it follows that the quantity 
$\left(\mu - m \right)/T$ is generally constant for nonrelativistic matter. 
It is well-known (see, e.g., \cite{Ehl,IsNot}) that $\mu - m$ is just the nonrelativistic chemical potential. 

This completes our general discussion of the thermodynamics of fluid 
particle creation in generalized equilibrium. 
We are now prepared to discuss the implications of this kind of particle production for the cosmological dynamics. 
While part of this discussion recalls earlier work one should keep in mind 
the much more general level of the present considerations. 
While the discussion in \cite{ZTP} relied on a very specific model for the creation process, the conclusions of this paper are completely model independent. 

\section{Particle production and cosmological dynamics}
The main advantage of a fluid approach to particle production is the possibility to calculate the backreaction of this process on the 
cosmological dynamics. 
As was shown in sections II and III this problem is equivalent to 
studying the dynamics of a bulk viscous fluid universe. 
Moreover, the effective bulk pressure that describes the production 
process in the equilibrium case is explicitly given in terms of 
$p$ and $\rho $, implying the dynamical laws (\ref{32}) and (\ref{33}). 
The expression $\partial \rho / \partial p$ in the latter equations is generally given by Eq. (\ref{50}). 
Explicit integration of (\ref{32}) and (\ref{33}) is possible in the 
limiting cases of radiation $p = nT$, $\rho = 3 nT$ and nonrelativistic 
matter 
$p = nT$, $\rho = nm + \frac{3}{2}nT$, $m \gg T$. 
Introducing a length scale $a$ according to 
\begin{equation}
\Theta \equiv  3 \frac{\dot{a}}{a}
\label{58}
\end{equation}
one recovers the familiar dependences 
\begin{equation}
n \propto a ^{-3}\ , 
\mbox{\ \ \ \ \ \ \ \ } 
\rho \propto a ^{-4} 
\mbox{\ \ \ \ \ \ \ \ \ \ \ \ \ }
\left(p = \rho /3 \right) 
\label{59}
\end{equation}
for radiation since $\Gamma $ as well as $\pi $ vanish in this case. 
For nonrelativistic matter on the other hand, one obtains
\begin{equation}
n \propto a ^{-3/2}\ , 
\mbox{\ \ \ \ \ \ \ \ \ \ }
\rho \propto a ^{-3/2} 
\mbox{\ \ \ \ \ \ \ \ \ \ \ } 
\left(m \gg T \right) \ .
\label{60}
\end{equation}
The temperature behaves according to Eq. (\ref{26}) 
for any equation of state. 
Since nonrelativistic matter is produced at a rate 
$\Gamma = \Theta /2$ (cf. Eq. (\ref{31})) its standard behaviour for 
the case of conserved particle number ( $n \propto a ^{-3}$, 
$T \propto a ^{-2}$, $\rho \propto a ^{-3}$) 
changes considerably. 
All the thermodynamic quantities decrease more slowly now since the decay of 
$n$, $T$ and $\rho $ due to the expansion is counteracted by corresponding production terms. 
The differences  in the behaviour of $n$, $T$ and $\rho$ acoording to the relations (\ref{60}) compared with the standard behaviour 
for $\Gamma = 0$ are 
consequences of the backreaction of the production process on the fluid dynamics. 
Also the expansion of the universe is modified in case matter in generalized equilibrium becomes dynamically dominating.  
In a homogeneous and isotropic universe the lenght scale $a$ coincides with 
the scale factor of the Robertson-Walker metric. 
Restricting ourselves to the spatially flat case, the scale factor obeys 
the equation 
\begin{equation}
3 \frac{\dot{a}^{2}}{a ^{2}} = \kappa \rho \ ,
\label{61}
\end{equation}
where $\kappa$ is Einstein's gravitational constant. 
For radiation we recover, of course, $a \propto t ^{1/2}$. 
Inserting, however, the energy density 
$\rho $ from Eq. (\ref{60}) into Eq. (\ref{61}) we find that 
the scale factor behaves such as
\begin{equation}
a \propto t ^{4/3} 
\mbox{\ \ \ \ \ \ \ \ \ \ \ \ \ \ }
\left(m \gg T \right) \ ,
\label{62}
\end{equation}
i.e., $\ddot{a} > 0$ instead of the familiar $a \propto t ^{2/3}$ with  
$\ddot{a} < 0$ for $\rho \propto a ^{-3}$ corresponding to $\Gamma = 0$. 
The production of massive particles in equilibrium implies accelerated expansion ( $\ddot{a}>0$), i.e., power law inflation. 
In other words, only in a power law inflationary universe an equilibrium 
such as discussed in this paper is possible. 

The radiation and nonrelativistic matter equations of state constitute the limiting cases of the fluid behavior. 
The backreaction is largest for $m \gg T$ and it vanishes for 
$m \ll T$, i.e., 
$m \rightarrow 0$. 
There exists an intermediate equation of state yielding $a \propto t$ 
with $\ddot{a} = 0$. 
The condition for $\ddot{a} = 0$ is 
\begin{equation}
\rho + 3 \left(p + \pi  \right) = 0 \ ,
\label{63}
\end{equation}
or, with $\pi $ from Eq. (\ref{30}), 
\begin{equation}
\frac{\partial{\rho }}{\partial{p}} = \frac{2 \rho }{\rho + p}\ .
\label{64}
\end{equation}
Inserting here Eq. (\ref{50}), we obtain the following cubic equation for 
$h /T \equiv  \left(\rho + p \right)/ \left(nT \right)$:
\begin{equation}
\left(\frac{h}{T} \right)^{3} - 5 \left(\frac{h}{T} \right)^{2} 
- \left[\left(\frac{m}{T} \right)^{2} - 3 \right]\frac{h}{T} 
- 2 = 0 \ .
\label{65}
\end{equation}
The quantity $h$ is the enthalpy per particle, given by \cite{Groot} 
\begin{equation}
h = m \frac{K _{3}\left(\frac{m}{T} \right)}
{K _{2}\left(\frac{m}{T} \right)}\ .
\label{66}
\end{equation}
$K _{2}$ and $K _{3}$ are Bessel functions of the second kind. 
The critical value $\left(m/T \right)_{cr}$ that corresponds to 
$\ddot{a} = 0$ turns out to be 
$\left(m/T \right)_{cr} \approx 9.55$. 
We have $\ddot{a} > 0$ for $\left(m/T \right) > \left(m/T \right)_{cr}$, 
while 
$\left(m/T \right) < \left(m/T \right)_{cr}$ corresponds to $\ddot{a} < 0$. 
In terms of a ``$\gamma $'' law, i.e., $p = \left(\gamma - 1 \right)\rho $, 
the corresponding critical value $\gamma _{cr}$ is 
$\gamma _{cr} \approx 1.09$. 
There is power law inflation for 
$1 \leq \gamma \leq \gamma _{cr}$. 
The case $\gamma = \gamma _{cr}$ corresponds to $\ddot{a} = 0$. 
We mention again that all the results concernig the backreaction of 
equilibrium particle production on the cosmological dynamics are quite 
general and do not rely on a specific creation model. 

In the following section we will establish relations between the results obtained so far and corresponding considerations on the level of 
relativistic kinetic theory. 
This will provide us with a microscopic justification of the 
phenomenological generalized equilibrium concept. 

\section{Kinetic theory and particle production}
\subsection{General relations}
For conserved particle numbers the fluid 
dynamics 
of a gas may be derived from Boltzmann's equation for the one-particle distribution function. 
One may address the question whether a 
corresponding microscopic justification also exists in the case of  
varying particle numbers. 
Previously \cite{TZP,ZTP} a modified Boltzmann equation was proposed 
in which an additional ``source term'' describes the change of the 
one-particle distribution function due to particle number nonconserving processes, supposedly of quantum origin. 

The one-particle distribution function $f = f \left(x,p \right)$ of a relativistic gas with varying particle number is supposed to obey the 
equation
\begin{equation}
L\left[f\right] \equiv 
p^{i}f,_{i} - \Gamma^{i}_{kl}p^{k}p^{l}
\frac{\partial f}{\partial
p^{i}}
 = C\left[f\right] + H\left(x, p\right)  \mbox{ , } 
\label{67}
\end{equation}
where $f\left(x, p\right) p^{k}n_{k}\mbox{d}\Sigma dP$ 
is the number of
particles whose world lines intersect the hypersurface element 
$n_{k}\mbox{d}\Sigma$ around $x$, having four-momenta in the range 
$\left(p, p + \mbox{d}p\right)$.  \\
$\mbox{d}P = A(p)\delta \left(p^{i}p_{i} + m^{2}\right) 
\mbox{d}P_{4}$
is the volume element on the mass shell 
$p^{i}p_{i} = -  m^{2} $
in the momentum space. 
$A(p) = 2$, if $p^{i}$ is future directed and 
$A(p) = 0$ otherwise; 
$\mbox{d}P_{4} = \sqrt{-g}\mbox{d}p^{0}\mbox{d}p^{1}
\mbox{d}p^{2}\mbox{d}p^{3}
$. \\
$C[f]$ is the Boltzmann collision term. 
Its specific structure discussed e.g. by 
Ehlers \cite{Ehl} will not 
be relevant for our considerations. 
Following Israel and Stewart \cite{IS} we shall only 
require that (i) $C$
is a local function of the distribution function, i.e., 
independent
of derivatives of $f$, (ii) $C$ is consistent with 
conservation of
four-momentum and number of particles, and (iii) $C$
yields a nonnegative expression for the entropy 
production and does
not vanish unless $f$ has the form of a local equilibrium
distribution (see Eq. (\ref{73}) below). 

The term $H(x,p)$ on the right-hand side of (\ref{67})  takes into 
account the fact
that the number of particles whose world lines intersect a 
given
hypersurface element within a certain range of momenta may 
additionally 
change due to creation or decay processes. 
On the level of classical kinetic theory we shall regard 
this term as 
a given
input quantity. 
Below we shall give an example for the possible functional 
structure
of $H(x, p)$. 

By the splitting of the right-hand side of Eq. (\ref{67}) into $C$ and 
$H(x, p)$ we
have separated the collisional from the creation (decay) 
events. 
In this setting collisions are not accompanied by creation or
annihilation processes. In other words, once created, 
the interactions between the particles 
are both energy-momentum and number preserving. 
For a vanishing $H(x, p)$ Eq. (\ref{67})  reduces to the 
familiar
Boltzmann equation (see, e.g., \cite{Stew,Ehl,IS,Groot}). 
 
The particle number flow four-vector 
$N^{i}$ and the energy momentum tensor $\tilde{T}^{ik}$ are
defined in a standard way (see, e.g., \cite{Ehl}) as 
\begin{equation}
N^{i} = \int \mbox{d}Pp^{i}f\left(x,p\right) \mbox{ , } 
\ \ \ 
\tilde{T}^{ik} = \int \mbox{d}P p^{i}p^{k}f\left(x,p\right) \mbox{ .} 
\label{68}
\end{equation}
While it will turn out that in the equilibrium case the 
first moment of $f$ 
in Eq. (\ref{68}) may be identified with 
the quantity (\ref{4}), we have used the 
symbol $N ^{i}$ immediately. 
The second moment of $f$ denoted by $\tilde{T}^{ik}$ in Eq. (\ref{68}) 
will not, however, coincide with the energy momentum tensor $T ^{ik}$ 
in Eq. (\ref{1}) with Eq. (\ref{2}). 
The integrals in (\ref{68}) and in the following 
are integrals over the entire mass shell 
$p^{i}p_{i} = - m^{2}$. 
The entropy flow vector $S^{a}$ is given by \cite{Ehl,IS} 
\begin{equation}
S^{a} = - \int p^{a}\left[
f\ln f - f\right]\mbox{d}P \mbox{ , }
\label{69}
\end{equation}
where we have restricted ourselves to the case of 
classical Maxwell-Boltzmann particles. 
 
Using the general relationship \cite{Stew}
\begin{equation}
\left[\int p^{a_{1}}....p^{a_{n}}p^{b}f \mbox{d}P\right]_{;b} 
= \int p^{a_{1}}...p^{a_{n}}L\left[f\right] \mbox{d}P 
\label{70}
\end{equation}
and Eq. (\ref{67}) we find 
\begin{equation}
N^{a}_{;a} = \int \left(C\left[f\right] + H\right) \mbox{d}P 
\mbox{ , } \ \ 
\tilde{T}^{ak}_{\ ;k} =  \int p^{a}\left(C\left[f\right] + H\right) 
\mbox{d}P
\mbox{ , } 
\label{71}
\end{equation}
and 
\begin{equation}
S^{a}_{;a} = - \int \ln f 
\left(C\left[f\right] + H\right) \mbox{d}P
\mbox{ .} 
\label{72}
\end{equation}
In collisional equilibrium which we
shall assume from
now on, $\ln f$ in Eq. 
(\ref{72}) 
is a linear combination of the collision invariants 
$1$ and $p^{a}$.  
The corresponding equilibrium distribution function 
becomes (see, e.g., \cite{Ehl}) 
\begin{equation}
f^{0}\left(x, p\right) = 
\exp{\left[\alpha + \beta_{a}p^{a}\right] } 
\mbox{ , }\label{73}
\end{equation}
where $\alpha = \alpha\left(x\right)$ and 
$\beta_{a}\left(x \right)$ is timelike. 
Inserting the function (\ref{73}) into Eq. (\ref{67}) one obtains 
\begin{equation}
\left[p^{a}\alpha_{,a} +
\beta_{\left(a;b\right)}p^{a}p^{b}\right]f^{0}  
=  H\left(x, p\right) 
\mbox{ .} \label{74}
\end{equation} 
\indent It is well known \cite{Ehl,IS} 
that for $H(x, p) = 0$ this equation, which characterizes 
the ``global equilibrium'', admits solutions
only for very special cases in which $\alpha = const$ and 
$\beta_{a}$ is a timelike Killing-vector. 
The only exception is the case $m = 0$, in which $\beta _{a}$ is only 
required to be a CKV in order to satisfy the equilibrium conditions. 
It will be our main objective in this section to explore whether a 
nonvanishing 
source term $H \left(x,p \right)$ admits Eq. (\ref{74}) to be 
fulfilled under the 
condition that $\beta _{a}$ is a CKV also for $m > 0$. 
With the equilibrium distribution funktion (\ref{73}) the balances 
(\ref{71}) 
reduce to    
\begin{equation}
N^{a}_{;a}=\int H\mbox{d}P \mbox{\ , \ \  } 
\tilde{T}^{ak}_{\ ;k}=\int p^{a}H\mbox{d}P \ .
\label{75}
\end{equation}
In collisional equilibrium there is entropy production only due to
the source term $H$. From Eq. (\ref{72}) we obtain 
\begin{equation}
S^{a}_{;a} = - \int H\left(x, p\right)
\ln f^{0} \mbox{d}P
\mbox{ , }\label{76}
\end{equation} 
implying $S^{a}_{;a} = - \alpha N^{a}_{;a} 
- \beta_{a}\tilde{T}^{ab}_{\ ;b}$. 
With $f$ replaced by $f^{0}$ in Eqs. 
(\ref{68}) and (\ref{69}), $N^{a}$, $\tilde{T}^{ab}$ and $S^{a}$ may be 
split with respect to the unique four-velocity $u^{a}$ according to 
\begin{equation}
N^{a} = nu^{a} \mbox{ , \ \ }
\tilde{T}^{ab} = \rho u^{a}u^{b} + p h^{ab} \mbox{ , \ \ }
S^{a} = nsu^{a} \mbox{  . }
\label{77}
\end{equation}
The exact integral expressions for $n$, $\rho$, $p$ and $s$ are given
by the formulae (177) - (180) in \cite{Ehl}. 

Using the first Eq. (\ref{77}) and defining
\begin{equation}
\Gamma  \equiv \frac{1}{n} \int H\left(x, p\right) \mbox{d}P 
\mbox{ , }
\label{78}
\end{equation}
the first Eq. (\ref{75}) coincides with Eq. (\ref{6}).  
It is obvious that $\Gamma$ is the particle production rate. 
Similarly, 
with the decomposition (\ref{77}) and the abbreviation
\begin{equation}
t^{a} \equiv - \int p^{a}H\left(x, p\right) \mbox{d}P \mbox{ , }
\label{80}
\end{equation} 
the second equation (\ref{75}) may be written as 
\begin{equation}
\tilde{T}^{ak}_{;k} + t ^{a} = 0 \ .
\label{81}
\end{equation}
For $t ^{a} \neq  0$ the energy momentum tensor $\tilde{T}^{ak}$ is not conserved. 
We are interested in the question whether it is possible to replace the left-hand side of the last equation, 
$\tilde{T}^{ak}_{;k} + t ^{a}$, by a locally conserved quantity, 
$T ^{ak}_{\ ;k}$, in such a way that the source term $t ^{a}$ in 
Eq. (\ref{81}) 
is mapped onto an effective viscous pressure $\pi $ of the energy momentum 
tensor $T ^{ak}$, which is to be identified with 
the corresponding quantity in Eq. (\ref{1}). 

The equations (\ref{81}) imply 
\begin{equation}
\dot{\rho } + \Theta \left(\rho + p \right) = u _{a} t ^{a}
\label{82}
\end{equation}
and
\begin{equation}
\left(\rho + p \right) \dot{u}_{a} + \nabla  _{a}p 
= - h _{ai}t ^{i }\ .
\label{83}
\end{equation}
Comparing these balances with relations (\ref{21}) 
(for $q ^{a} = \sigma _{ab} = 0$) 
and (\ref{39}) suggests the 
tentative identifications
\begin{equation}
\Theta \pi \equiv  - u _{a} t ^{a}
\label{84}
\end{equation}
as well as
\begin{equation}
\pi \dot{u}_{a} + \nabla  _{a}\pi = h _{ai}t ^{i} \ .
\label{85}
\end{equation}
In order to check the consistency of this interpretation it is necessary 
to specify the source term $H \left(x,p \right)$ which describes the 
variation of the one-particle distribution function due to the change 
in the number of gas particles. 
In the following subsection we will generalize previously used 
expressions for 
$H \left(x,p \right)$. 

\subsection{Effective rate approximation}
The quantity $H\left(x, p\right)$ is an
input quantity on the level of classical kinetic theory. It is 
supposed to represent the net effect of certain quantum
processes with variable particle numbers (see, e.g., \cite{BiDa,Ford}). 
Also internal interactions within the cosmological fluid may give rise to 
an increase in the overall number of particles \cite{ZMNRAS,ZPM}. 
Following earlier considerations \cite{TZP,ZTP} we assume 
$H\left(x, p\right)$ to depend linearly on the equilibrium distribution 
function (\ref{73}), i.e., 
\begin{equation}
H\left(x, p\right) = \zeta\left(x, p\right) f^{0}\left(x, p\right) 
\mbox{ .}
\label{86}
\end{equation}
We also assume that $\zeta \left(x,p \right)$ depends only linearly on the momenta $p ^{a}$. This ensures that the source terms $\Gamma $ 
and $t ^{a}$ in the balances (\ref{6}) and (\ref{81}) do not depend on 
moments of the distribution function higher than the second one. 
The four-momenta $p ^{a}$ may be decomposed into 
\begin{equation}
p ^{a} = E u ^{a} + \lambda e ^{a} \ ,
\label{87}
\end{equation}
where $e ^{a}$ is a unit spatial vector, i.e., $e ^{a}e _{a} = 1$, 
$e ^{a}u _{a} = 0$. 
Consequently, one has $E = - u _{a}p ^{a}$ and $\lambda = e _{a}p ^{a}$ 
and the 
mass shell condition $p ^{a}p _{a} = - m ^{2}$ is equivalent to 
$\lambda ^{2} = E ^{2} - m ^{2}$. 
Moreover, $h _{ab}p ^{a}p ^{b} = \lambda ^{2}$ is valid. 
These properties suggest assuming $\zeta $ to consist additively of 
three terms: A term $\nu$ which does not depend on $p ^{a}$ at all, 
a term proportional to $E = - u _{a}p ^{a}$, and, finally, a term 
proportional to $\lambda = e _{a}p ^{a}$: 
\begin{equation}
\zeta \left(x,p \right) =  \nu\left(x\right)  
+ \frac{E}{\tau\left(x\right)} 
+ \frac{\lambda }{\sigma \left(x\right)}\ .
\label{88}
\end{equation}
This ansatz for $\zeta $ represents the most general structure compatible 
with a linear dependence of $\zeta $ on $p ^{a}$. 
The rate of change of the (equilibrium) distribution function is then determined by the three spacetime functions 
$\nu$, $\tau $, and $\sigma $. 
Different from earlier considerations \cite{TZP,ZTP} we have introduced 
here a term proportional to $\lambda = e _{a}p ^{a} = \sqrt{h _{ab}p ^{a}p ^{b}}$. 
As we shall see below this more general structure will remove some 
unpleasant properties of the model studied in \cite{TZP,ZTP}. 

With the relations (\ref{86}) and (\ref{88}) for $H \left(x,p \right)$ 
the source terms $n \Gamma $ and $t ^{a}$ in Eqs. (\ref{78}) and 
(\ref{80}) are given by
\begin{equation}
n\Gamma  =  \nu\left(x\right)M\left(x\right) 
- \frac{u_{a}}{\tau}N^{a}
= \nu M + \frac{n}{\tau }  
\mbox{ , }\label{89}
\end{equation}
where  $M$ is the zeroth moment of the distribution function,  
$M \equiv \int \mbox{d}Pf\left(x, p\right)$, and 
\begin{equation}
t^{a} = - \nu N ^{a} + \frac{u_{b}T ^{ab}}{\tau \left(x \right)} 
- \frac{e _{b}T ^{ab}}{\sigma \left(x \right)}
\mbox{ , }
\label{90}
\end{equation}
respectively. 
The last relations imply 
\begin{equation}
u _{i}t ^{i} = \nu n + \frac{\rho }{\tau }\ , \ \ \ \ \ \ \ \ \ 
h _{ci}t ^{i} = - \frac{e _{c}}{\sigma }p \ .
\label{91}
\end{equation}
Via the functions $\nu$, $\tau $, and $\sigma $ the source terms 
$\Gamma $, $u _{i}t ^{i}$, and $h _{ci}t ^{i}$ are coupled to the fluid quantities $n$, $\rho $, and $p$, representing the first and second moments 
of the distribution function. 
Because of the restriction to a linear dependence of $\zeta $ on 
$p ^{a}$ in Eq. (\ref{88}), no higher-order moments appear in 
the expressions 
(\ref{89})  and (\ref{90}). 
Using the structure (\ref{86}) with 
the ansatz (\ref{88}) for the source term 
$H \left(x,p \right)$ in Eq. (\ref{74}), the latter reduces to the 
generalized equilibrium condition 
\begin{equation}
p^{a}\alpha_{,a}+\beta_{(a;b)}p^{a}p^{b} = \nu + \frac{E}{\tau} 
+ \frac{\lambda }{\sigma }\ .
\label{92}
\end{equation}
With the decomposition (\ref{87})  condition (\ref{92}) is satisfied for 
\begin{equation}
\dot{\alpha}=\frac{1}{\tau}\ ,\ \ \ \ 
e ^{a}\nabla  _{a}\alpha = \frac{1}{\sigma }\ ,
\label{93}
\end{equation} 
and
\begin{equation}
\beta_{(a;b)}= \phi (x)g_{ab}\ , 
\label{94}
\end{equation}
with $\phi  = - \nu /m ^{2}$. Together with Eq. (\ref{17}) this implies 
\begin{equation}
\frac{\nu}{m ^{2}} = - \frac{1}{3}\frac{\Theta }{T}\ .
\label{95}
\end{equation} 
Using the identification $\beta _{a} = u _{a}/T$ we have recovered the CKV property (\ref{15}) as a (generalized) equilibrium condition for the distribution 
function (\ref{73}). 
According to the relations (\ref{93}) the function $\alpha$ may vary 
both in time and in space. 
The possibility of a spatial variation is due to the last term in the expression (\ref{88}) and was not taken into account in \cite{TZP,ZTP}. 
As we will see below, a spatially varying $\alpha$ is, however, essential 
to establish the full correspondence between the microscopic and macroscopic levels of description. 
Identifying $\alpha$ with $\mu /T$ of the phenomenological theory, 
the relation 
(\ref{76}) may be written as
\begin{equation}
S ^{a}_{;a} = - \frac{\mu }{T}n \Gamma 
- \frac{u _{a}t ^{a}}{T}\ .
\label{96}
\end{equation}
Using here Eq. (\ref{48}) we obtain 
\begin{equation}
S ^{a}_{;a} - ns \Gamma = - \frac{\rho + p}{T}\Gamma 
+ \frac{u _{a}t ^{a}}{T }\ .
\label{97}
\end{equation}
Introducing the abbreviation (\ref{84}) for $\pi $ into the last relation 
and taking into account the conditions (\ref{25}) we 
recover Eq. (\ref{24}). 
For $q ^{a} = 0$ 
the right-hand side of Eq. (\ref{97}) coincides with $n T \dot{s}$ 
(cf. Eq. (\ref{23})). 
With $\alpha = \mu /T$, the first condition 
(\ref{93}) may be combined with Eq. (\ref{37}) to yield 
\begin{equation}
\frac{1}{\tau } = \frac{1}{3}\left(\frac{\rho }{p} 
- \frac{\partial{\rho }}{\partial{p}}  \right)\Theta \ .
\label{98}
\end{equation}
On the other hand, $\tau ^{-1}$ and $\Gamma $ are related by 
Eq. (\ref{89}). 
It follows with Eq. (\ref{29}) that the function $\nu$ is given by 
\begin{equation}
\nu = \frac{n}{M}\left(1 - \frac{\rho }{3p} \right)\Theta \ .
\label{99}
\end{equation}
The second equation (\ref{93}) together with Eq. (\ref{43}) fixes the 
function $\sigma $:
\begin{equation}
\frac{1}{\sigma } = \sqrt{\nabla  ^{a}
\left(\frac{\mu }{T}\right) \nabla _{a}
\left(\frac{\mu }{T}\right)}  
= \sqrt{\left[\frac{\nabla  ^{a}n}{n} 
- \frac{\rho }{p}\frac{\nabla ^{a}T}{T} \right]
\left[\frac{\nabla  _{a}n}{n} 
- \frac{\rho }{p}\frac{\nabla  _{a}T}{T} \right]
} \ .
\label{100}
\end{equation}
An alternative expression for $1/ \sigma $ maybe obtained by using 
formula (\ref{52}) for $\nabla  _{a}\left(\mu /T \right)$. 

With the first relation (\ref{91}) and the expressions (\ref{99}) for 
$\nu$ and (\ref{98}) for $\tau ^{-1}$ the energy balance (\ref{82}) 
is consistent with the balance $\dot{\rho } + 
\Theta \left(\rho + p + \pi  \right) = 0$ of the 
phenomenological approach with $\pi $ given by 
relation (\ref{30}), equivalent to 
Eq. (\ref{84}). 
To check the corresponding property for the momentum balance one has 
to show the equivalence between Eqs. (\ref{83}) and 
(\ref{39}). 
The identification $\alpha = \mu /T$ allows us to combine 
Eq. (\ref{40}) with the second equation (\ref{93}). 
Together with the second relation (\ref{91}) this justifies relation 
(\ref{85}). 
Consequently, the balances (\ref{81}) with a nonconserved energy-momentum tensor 
$\tilde{T}^{ak}$ given by $\tilde{T}^{ab} = \rho u ^{a}u ^{b} 
+ p h ^{ab}$ are entirely equivalent to the local conservation laws 
$T ^{ak}_{\ ;k} = 0$ for the energy-momentum tensor 
$T^{ak} = \rho u ^{a}u ^{k} 
+ \left(p + \pi  \right) h ^{ak}$. 
The source terms $t ^{a}$ in the balances (\ref{81}) have consistently 
been mapped onto the quantity $\pi $ in the conserved energy-momentum 
tensor 
$T ^{ak}$. 
This completes our microscopic derivation of generalized equilibrium 
including the equivalence between 
particle production processes and effective viscous pressures.  

\section{Summary}
The ``generalized'' equilibrium concept introduced in this paper 
allows one to characterize cosmological fluids with growing number of particles. 
This equilibrium relies on the conformal Killing vector property of 
$u _{i}/T$ and implies nonvanishing entropy production which, however, is entirely due to an 
enlargement of the phase space of the system but not to dissipative processes. 
For a gas the ``generalized'' equilibrium 
may be derived from relativistic kinetic theory.  
The corresponding gas particles are governed 
by an equilibrium distribution function although 
their number is generally increasing.  
In the limit of relativistic particles with equation of state 
$p = \rho /3$ the ``generalized'' equilibrium reduces to the well-known ``global'' equilibrium of standard relativistic thermodynamics. 
Nonrelativistic particles may be in generalized equilibrium 
only in a power law inflationary universe. 
The latter statement confirms and generalizes earlier results. 
It was shown here to be model independent.

\acknowledgments
This paper was supported by the Deutsche Forschungsgemeinschaft,  
the CNRS and the NATO (grant CRG940598). 
Discussions with Diego Pav\'{o}n, J\'{e}r\^{o}me Gariel and 
Alexander Balakin  are gratefully acknowledged. 
I thank the members of the Laboratoire de Gravitation et Cosmologie Relativistes, Universit\'{e} Pierre et Marie Curie, Paris
for warm hospitality.

\end{document}